\documentclass[aps,showpacs,amsfonts,amsmath,prb,endfloats*,preprint]{revtex4}

\usepackage{graphicx}  
\usepackage{array}

\newcolumntype{x}[1]{%
>{\centering\hspace{0pt}}p{#1}}%

\usepackage{color}
 
\let\origcite\cite
\def\cite{\unskip\origcite}

\begin{document}
\title{Micelle shape transitions in block copolymer/homopolymer blends: comparison of self-consistent field theory with experiment}
\date\today
\author{M.~J.~Greenall}

\affiliation{School of Physics and Astronomy, University of Leeds, Leeds LS2 9JT, U.K.}
\author{D.~M.~A.~Buzza}

\affiliation{Department of Physics, The University of Hull, Cottingham Road, Hull HU6 7RX, U.K.}
\author{T.~C.~B.~McLeish}

\affiliation{School of Physics and Astronomy, University of Leeds, Leeds LS2 9JT, U.K. and Department of Physics, Durham University, South Road, Durham DH1 3LE, U.K.}
\begin{abstract}
Diblock copolymers blended with homopolymer may self-assemble into spherical, cylindrical or lamellar aggregates. Transitions between these structures may be driven by varying the homopolymer molecular weight or the molecular weight or composition of the diblock. Using self-consistent field theory (SCFT), we reproduce these effects. Our results are compared with X-ray scattering and transmission electron microscopy measurements by Kinning, Winey and Thomas and good agreement is found, although the tendency to form cylindrical and lamellar structures is sometimes overestimated due to our neglect of edge effects due to the  finite size of these aggregates. Our results demonstrate that self-consistent field theory can provide detailed information on the self-assembly of isolated block copolymer aggregates. 
\end{abstract}

\maketitle

\section{Introduction}

Block copolymers are formed from two or more types of monomer, which are linked in such a way that monomers of a given type are grouped together in long intervals, or blocks \cite{hamley_book}. These materials not only have engineering applications, for example in lithography \cite{zschech}, but can also mimic biological systems \cite{jain_bates}. The utility of block copolymers arises from their ability to {\em self-assemble} into a range of structures. A striking example of this behavior is seen when the block copolymers are dissolved in a solvent such as a liquid or a homopolymer. Consider a simple block copolymer with two sections (a {\em diblock} copolymer), one of which is solvophobic, or incompatible with the solvent. At high enough copolymer concentration, the solvophobic segments of the polymer will cluster together in order to minimize their contact with the solvent. A wide variety of structures can be formed in this way \cite{safran_book,jain_bates}, including spheres and cylinders of copolymer known as {\em micelles} and hollow pockets ({\em vesicles}). Both micelles and vesicles may be used to encapsulate active chemicals such as drugs \cite{kim}. Although the basic principle behind the self-assembly of block copolymers can be easily explained, making quantitative predictions about which structure will be formed by a given system is a much more difficult problem. This depends sensitively on many different factors \cite{battaglia_ryan}, particularly the structure of the copolymer molecules. Designing block copolymers that will self-assemble into the structure required by a particular application can hence be difficult, and there is a clear need for theoretical work to provide guidance for experiment.

A theory with strong potential for furthering our understanding of the self-assembly of block copolymers in solution is {\em self-consistent field theory} (SCFT), a mean-field theory of an ensemble of flexible polymers. This theory has had much success in modeling melts and blends of block copolymers \cite{matsen_book}. However, much previous research in the field has focused on the formation of periodic structures \cite{maniadis}, with the SCFT equations often being solved in Fourier space. In this paper, we present a detailed comparison of {\em real-space} SCFT with experiment for local, isolated aggregates of block copolymers. We concentrate on the formation of spherical micelles, cylindrical micelles, and flat bilayers in blends of diblock copolymer and homopolymer and predict which of these structures will be formed for a given blend. This system is well suited to provide a test of SCFT, for a number of reasons. Firstly, detailed experimental data on the formation of different aggregates are available \cite{kinning_winey_thomas}. In addition, and in contrast to the situation in aqueous solutions, the interactions between the different types of polymer are well described by the Flory $\chi$ parameter \cite{jones_book}. This parameter, and the other quantities (such as molar volumes) needed as input to SCFT, are readily obtained from the literature \cite{polymer_handbook}.

We compare our predictions with the X-ray scattering and transmission electron microscopy experiments of Kinning, Winey and Thomas \cite{kinning_winey_thomas}, which study poly(styrene-butadiene) diblocks in homopolystyrene. To the best of our knowledge, such data have not been modeled in detail, using experimentally-determined polymer architectures and molecular weights, with SCFT. Although a significant amount of research has been carried out on the self-consistent field theory of micelle formation, this largely considers diblocks dissolved in {\em monomer} solvent \cite{leermakers,leermakers_scheutjens-shape,linse}.

The details of the polystyrene/polybutadiene (PS/PB) system are presented in Sec.\ \ref{system_details}, along with the basic phenomenology of shape transitions. Section \ref{scft} provides a brief overview of the self-consistent field theory of a copolymer/homopolymer blend, and introduces the numerical methods used to solve the SCFT equations. We then present and discuss our results (Sec.\ \ref{results}) and the conclusions are given in Sec. \ref{conclusions}.

\section{Details of system and phenomenology of shape transitions}\label{system_details}

Kinning, Winey and Thomas \cite{kinning_winey_thomas} carried out X-ray and transmission electron microscopy measurements on blends of poly(styrene-butadiene) diblock copolymer and polystyrene homopolymer. In order to determine the effects of the molecular weight of the polymers and the relative amounts of styrene and butadiene in the diblocks on the shape of the aggregates formed, several samples were studied. We follow the notation used in this paper \cite{kinning_winey_thomas} when labeling these blends. For example, a diblock of polystyrene molar weight 10 kg/mol and polybutadiene molar weight 65 kg/mol is referred to as SB 10/65. Homopolystyrene of molar weight 2.1 kg/mol is labeled 2100 PS. The numbers used for labeling purposes can be quite rough: the precise molecular weights used in our calculations can be found in the original paper \cite{kinning_winey_thomas}.

We now introduce the quantities by which the polymer samples are characterized: their specific volumes, the root mean square end-to-end distances of the polymer molecules and the {\em interaction energy density} of polystyrene and polybutadiene.

Measurements of the specific volume of PB as a function of temperature (in degrees Celsius) can be fitted empirically by \cite{polymer_handbook}
\begin{equation}
V_\text{PB}(\text{cm}^3/\text{g})\approx 1.0968 + (8.24\times 10^{-4})T
\label{V_PB}
\end{equation}
whilst the corresponding formula for PS is \cite{richardson_savill}
\begin{equation}
V_\text{PS}(\text{cm}^3/\text{g})\approx 0.9217 + (5.412\times 10^{-4})T + (1.678\times 10^{-7})T^2
\label{V_PS}
\end{equation}
The temperature $T$ is $115^\circ\text{C}$ in all experiments considered here.

The SCFT equations are usually written in terms of the volumes of individual molecules $v_i$, with $i$ representing PB or PS. These are calculated from Eqns \ref{V_PB} and \ref{V_PS} by \cite{kinning}
\begin{equation}
v_i({\text{\AA}^3})=M_i(\text{g/mol})\times V_i(\text{cm}^3/\text{g})/0.602
\label{molecular_volume}
\end{equation}
where $M_i$ is the molar weight of polymer $i$.

The root-mean-square end-to-end distances of the polymer molecules are given, also empirically, by \cite{kinning}
\begin{eqnarray}
\langle R^2_\text{PB}\rangle^{1/2}(\text{\AA}) & \approx & 0.93M_\text{PB}^{1/2}\nonumber\\
\langle R^2_\text{PS}\rangle^{1/2}(\text{\AA}) & \approx & 0.70M_\text{PS}^{1/2}
\label{rms}
\end{eqnarray}
where the molar weights $M_i$ are in g/mol. The experimental work considered here \cite{kinning_winey_thomas,kinning} describes the strength of the interaction between polystyrene and polybutadiene using the {\em interaction energy density} $\Lambda$ \cite{roe}, rather than the more usual $\chi$ parameter. The energy of this interaction is given by
\begin{equation}
\Lambda\int\mathrm{d}\mathbf{r}\,\phi_\text{PS}(\mathbf{r})\phi_\text{PB}(\mathbf{r})
\label{Lambda_interaction}
\end{equation}
where the $\phi_i(\mathbf{r})$ terms are the local volume fractions of the two polymer species $i$ at position $\mathbf{r}$.

For reference, the interaction energy density $\Lambda$ is related to the $\chi$ parameter by \cite{roe}
\begin{equation}
\Lambda = \chi(V_\text{ref}/k_\text{B}T)^{-1}
\label{Lambda_chi}
\end{equation}
where $k_\text{B}$ is Boltzmann's constant. $V_\text{ref}$ is a reference volume, such as the average of the repeat unit volumes of the two polymers \cite{matsen_book}. The advantage of working in terms of $\Lambda$ rather than $\chi$ is that it avoids the introduction of this arbitrary reference volume \cite{roe}.

The temperature dependence of the polystyrene-polybutadiene interaction energy density is given, again empirically, by \cite{roe_zin}
\begin{equation}
\Lambda(\text{cal}/\text{cm}^3)=A-B(T-150^\circ\text{C})
\label{Lambda_T}
\end{equation}
where $A=0.718\pm 0.051$ and $B=0.0021\pm 0.00045$.

Depending on the molecular weights of the different components of the blend and the ratio of PB to PS in the copolymer, the system was found to form spherical or cylindrical micelles or bilayers \cite{kinning_winey_thomas}. We now present a brief discussion of the factors that affect which of these is most likely to form. Consider a melt of symmetric copolymers, such as SB 20/20, with no homopolystyrene. This will form a periodic, lamellar structure, since the layers have no natural curvature. However, if polystyrene homopolymer is added, this will mix preferentially with the polystyrene segments of the copolymer, leading the PS/PB interface to become curved and cylindrical or spherical micelles to form \cite{kinning_winey_thomas}. The degree of swelling determines which morphology will be observed: if a large amount of homopolymer penetrates the polystyrene corona, the interface between the two species will have a high curvature and spherical micelles will be favored. When less swelling of the corona takes place, cylindrical micelles will form.

It is also possible to form (nearly) planar bilayers in a blend of diblock copolymer and homopolymer if the diblock polymer is strongly asymmetric, with the polybutadiene core block being several times heavier than the polystyrene corona block \cite{kinning_winey_thomas}. In this case, swelling of the polystyrene by the homopolystyrene solvent may balance the effect of the larger polybutadiene block and lead to roughly equal effective volume fractions for the two species. Bilayers which are too large have a strong tendency to form vesicles, to eliminate the energy cost of forming edges. However, in this paper we will only consider the formation of infinite bilayers and will delay the detailed discussion of vesicle formation to a future study.

\section{Self-consistent field theory of polymer statics}\label{scft}

Self-consistent field theory (SCFT) \cite{edwards} is an equilibrium, mean-field theory of a melt or blend of polymers. The description of the polymers is coarse-grained: the configuration of an individual polymer molecule is taken to be a random walk in space $\mathbf{r}_\alpha(s)$, where $s$ is a curve parameter specifying the position along the molecule. The interactions between polymers are modeled by assuming that the blend is incompressible and introducing a contact potential between molecules of different species. As discussed in Sec.\ \ref{system_details}, the strength of this potential is specified by the interaction energy density $\Lambda$.

Simulation of the system described above for a realistically large number of molecules would require a tremendous amount of computing power. SCFT lowers the computational requirements sharply by first reducing the $N$-body problem of modeling an ensemble of $N$ polymers of $i$ different species to $i$ $1$-body problems, and then introducing mean-field approximations to make these computationally tractable. The first step in this procedure is to view each molecule as being acted on by a field produced by all other molecules in the blend \cite{matsen_book}. This transforms the $N$-body problem into $N$ $1$-body problems. Since we wish to compute the partition sum over all configurations of the system, all molecules of a given species may be treated as equivalent. Therefore, we need only introduce one field $W_i(\mathbf{r})$ for each species $i$ and have only to solve $i$ $1$-body problems. Note that no approximation has yet been made -- the complexity of the system is now contained in the fields $W_i(\mathbf{r})$, which are yet to be calculated. The success of SCFT arises from the fact that approximations may be found more easily for the fields than for the original formulation of the problem. 

We now outline the rest of the derivation of SCFT for the specific case of our diblock copolymer/homopolymer blend. As discussed above, we introduce fields $W_\text{PS}$, $W_\text{PB}$ and $W_\text{hPS}$ acting on the polystyrene blocks, polybutadiene blocks and homopolystyrene respectively. This partition sum is thus converted into a functional integral over fields, with the original Hamiltonian replaced by an effective Hamiltonian $H$. By adapting standard derivation \cite{matsen_book}, we find that $H$ is given by
\begin{eqnarray}
\lefteqn{\frac{H}{k_\text{B}T} = \frac{\Lambda}{k_\text{B}T}\int\mathrm{d}\mathbf{r}\,[\Phi_\text{PS}(\mathbf{r})+\Phi_\text{hPS}(\mathbf{r})][1-\Phi_\text{PS}(\mathbf{r})-\Phi_\text{hPS}(\mathbf{r})]}\nonumber\\ & &
-\frac{1}{v_\text{PS}+v_\text{PB}}\int\mathrm{d}\mathbf{r}\,\{W_\text{PS}(\mathbf{r})\Phi_\text{PS}(\mathbf{r})+W_\text{PB}(\mathbf{r})[1-\Phi_\text{PS}(\mathbf{r})-\Phi_\text{hPS}(\mathbf{r})]\}\nonumber\\ & &
-\frac{1}{v_\text{hPS}}\int\mathrm{d}\mathbf{r}\,W_\text{hPS}(\mathbf{r})\Phi_\text{hPS}(\mathbf{r})+\frac{\overline{\phi}_\text{hPS}V}{v_\text{hPS}}\left[\ln \left(\frac{\overline{\phi}_\text{hPS}V}{Q_\text{hPS}}\right)-1\right]\nonumber\\ & &
+\frac{(\overline{\phi}_\text{PS}+\overline{\phi}_\text{PB})V}{v_\text{PS}+v_\text{PB}}\left[\ln \left(\frac{(\overline{\phi}_\text{PS}+\overline{\phi}_\text{PB})V}{Q_\text{PS,PB}}\right)-1\right]
\label{hamiltonian}
\end{eqnarray}
where the $\Phi_i(\mathbf{r})$ are the local volume fractions of the various polymer species $i$ ($i=\text{PS}$, PB or hPS) and $V$ is the volume of the system. The mean volume fraction of species $i$ is given by $\overline{\phi}_i$ and $v_i$ is the volume of an individual molecule of this species. The first term gives the energy of the interaction between the different polymer species. In this term, and throughout, the polybutadiene block volume fraction $\Phi_\text{PB}(\mathbf{r})$ is replaced by $1-\Phi_\text{PS}(\mathbf{r})-\Phi_\text{hPS}(\mathbf{r})$, since the blend is incompressible and the local volume fractions must add to $1$ at every point. The terms involving the $W_i(\mathbf{r})$ all arise from the unit operators that are inserted into the partition function to convert the partition sum into an integral over fields. In the penultimate term, $Q_\text{hPS}$ is the partition function of a single homopolymer molecule acted on by the field $W_\text{hPS}(\mathrm{r})$. Similarly, $Q_\text{PS,PB}$ is the partition function of a single copolymer molecule subject to the fields $W_\text{PS}(\mathrm{r})$ and $W_\text{PB}(\mathrm{r})$. These are given by (again adapting standard derivations \cite{matsen_book})
\begin{eqnarray}
Q_\text{hPS}[W_\text{hPS}]&=&\int\mathrm{d}\mathbf{r}\,q_\text{hPS}(\mathbf{r},s)q_\text{hPS}^\dagger(\mathbf{r},s)\nonumber\\
Q_\text{PS,PB}[W_\text{PS},W_\text{PB}]&=&\int\mathrm{d}\mathbf{r}\,q_\text{PS,PB}(\mathbf{r},s)q_\text{PS,PB}^\dagger(\mathbf{r},s)
\label{single_chain_partition}
\end{eqnarray}
where the $q$ and $q^\dagger$ terms are single chain propagators \cite{matsen_book}. These satisfy diffusion equations with a field term, reflecting the fact that the polymer molecules are modeled as random walks acted on by an external field that incorporates their interactions with the rest of the melt. In the case of the homopolymer, we have
\begin{equation}
\frac{\partial}{\partial s}q_\text{hPS}(\mathbf{r},s)=\left[\frac{1}{6}\langle R^2_\text{hPS}\rangle\nabla^2-W_\text{hPS}(\mathbf{r})\right]q_\text{hPS}(\mathbf{r},s)
\label{diffusion}
\end{equation}
with initial condition $q_\text{hPS}(\mathbf{r},0)=1$. The curve parameter $s$ runs from $0$ to $1$ along the length of the molecule.

The case of the copolymer is slightly more complicated, since we must take into account the two different polymer species. This means that the diffusion equation for the copolymer must be solved with the field $W_i(\mathbf{r})$ and the prefactor of the $\nabla^2q$ term appropriate to each of the two sections of the copolymer \cite{fredrickson_book}, so that
\begin{eqnarray}
\frac{\partial}{\partial s}q_\text{PS,PB}(\mathbf{r},s)&=&\left[\frac{1}{6}\frac{\langle R^2_\text{PS}\rangle}{f}\nabla^2-W_\text{PS}(\mathbf{r})\right]q_\text{PS,PB}(\mathbf{r},s)\quad 0\le s \le f\nonumber\\
\frac{\partial}{\partial s}q_\text{PS,PB}(\mathbf{r},s)&=&\left[\frac{1}{6}\frac{\langle R^2_\text{PB}\rangle}{1-f}\nabla^2-W_\text{PB}(\mathbf{r})\right]q_\text{PS,PB}(\mathbf{r},s)\quad f < s \le 1
\label{copdiffusion}
\end{eqnarray}
with initial condition $q_\text{PS,PB}(\mathbf{r},0)=1$. $f$ is the volume fraction of polystyrene in the copolymer. Equations \ref{copdiffusion} have been written in such a way that we can use the empirical forms (Eqn.\ \ref{rms}) for the root-mean-square end-to-end distances. This, along with the fact that the curve parameter $s$ is chosen to run from $0$ to $1$, means that they take a slightly different form (with extra factors of $1/f$ and $1/(1-f)$) from corresponding equations elsewhere in the SCFT literature \cite{fredrickson_book}.

Until now, all steps have been exact. We now introduce the main approximation of SCFT. This consists of minimizing the effective Hamiltonian $H$ with respect to all fields $W_i(\mathbf{r})$ and all densities $\Phi_i(\mathbf{r})$, yielding a saddle-point approximation to the system partition function $Z$. The approximation is most effective when the polymers are long and fluctuations are weak. Here, it successfully isolates the dominant contribution to the partition function, and SCFT agrees well with experiment \cite{bates_fredrickson_rev}.

The minimisation of $F$ leads to a set of simultaneous equations relating the values of the fields and densities at the minimum, which we denote by lower-case letters $\phi_i(\mathbf{r})$ and $w_i(\mathbf{r})$. We find that
\begin{eqnarray}
1&=&\phi_\text{PS}(\mathbf{r})+\phi_\text{PB}(\mathbf{r})+\phi_\text{hPS}(\mathbf{r})\nonumber\\
\frac{1}{v_\text{PS}+v_\text{PB}}\left[w_\text{PS}(\mathbf{r})-w_\text{PB}(\mathbf{r})\right]&=&\frac{2\Lambda}{k_\text{B}T}\left[\overline{\phi}_\text{PS}+\overline{\phi}_\text{hPS}-\phi_\text{PS}(\mathbf{r})-\phi_\text{hPS}(\mathbf{r})\right]\nonumber\\
w_\text{hPS}&=&\frac{v_\text{hPS}}{v_\text{PS}+v_\text{PB}}w_\text{PS}(\mathbf{r})
\label{SCFT_equations}
\end{eqnarray}
where $\overline{\phi}_i$ is the (mean) volume fraction of species $i$. The first of these equations imposes the incompressibility of the melt. The densities are calculated from the propagators (see Eqn.\ \ref{diffusion}) according to \cite{matsen_book}
\begin{eqnarray}
\phi_\text{hPS}(\mathbf{r})&=&\frac{V\overline{\phi}_\text{hPS}}{Q_\text{hPS}[w_\text{hPS}]}\int^1_0\mathrm{d}s\, q_\text{hPS}(\mathbf{r},s)q_\text{hPS}^\dagger(\mathbf{r},s)\nonumber\\
\phi_\text{PS}(\mathbf{r})&=&\frac{V(\overline{\phi}_\text{PS}+\overline{\phi}_\text{PB})}{Q_\text{PS,PB}[w_\text{PS},w_\text{PB}]}\int^f_0\mathrm{d}s\, q_\text{PS,PB}(\mathbf{r},s)q_\text{PS,PB}^\dagger(\mathbf{r},s)\nonumber\\
\phi_\text{PB}(\mathbf{r})&=&\frac{V(\overline{\phi}_\text{PS}+\overline{\phi}_\text{PB})}{Q_\text{PS,PB}[w_\text{PS},w_\text{PB}]}\int^1_f\mathrm{d}s\, q_\text{PS,PB}(\mathbf{r},s)q_\text{PS,PB}^\dagger(\mathbf{r},s)
\label{density}
\end{eqnarray}
Note that, when calculating the copolymer densities, the integration limits are set to give the correct proportions of PS and PB.

To assess which of the possible structures is likely to form, we need to calculate their free energies, or, more accurately, their free energy densities. The SCFT approximation to the free energy density is obtained by substituting the self-consistent field equations (\ref{SCFT_equations}) into the effective Hamiltonian Eqn.\ \ref{hamiltonian}. This yields
\begin{eqnarray}
\lefteqn{\frac{A}{Vk_\text{B}T}=\frac{\overline{\phi}_\text{PS}+\overline{\phi}_\text{PB}}{v_\text{PS}+v_\text{PB}}\left(\ln(\overline{\phi}_\text{PS}+\overline{\phi}_\text{PB})-1\right)+\frac{\overline{\phi}_\text{hPS}}{v_\text{hPS}}\left(\ln \overline{\phi}_\text{hPS}-1\right)}\nonumber\\ & &
+\frac{\Lambda}{k_\text{B}T}(\overline{\phi}_\text{PS}+\overline{\phi}_\text{hPS})(1-\overline{\phi}_\text{PS}-\overline{\phi}_\text{hPS})\nonumber\\ & &
+\frac{\Lambda}{Vk_\text{B}T}\int\mathrm{d}\mathbf{r}(\phi_\text{PS}(\mathbf{r})+\phi_\text{hPS}(\mathbf{r})-\overline{\phi}_\text{PS}-\overline{\phi}_\text{hPS})(\phi_\text{PS}(\mathbf{r})+\phi_\text{hPS}(\mathbf{r})-\overline{\phi}_\text{PS}-\overline{\phi}_\text{hPS})\nonumber\\ & &
-\frac{\overline{\phi}_\text{PS}+\overline{\phi}_\text{PB}}{v_\text{PS}+v_\text{PB}}\ln\left(\frac{Q_\text{PS,PB}}{V}\right)-\frac{\overline{\phi}_\text{hPS}}{v_\text{hPS}}\ln\left(\frac{Q_\text{PS,PB}}{V}\right)
\label{SCFT_FE}
\end{eqnarray}
The densities are calculated according to Eqns\ \ref{density} and the single-chain partition functions according to Eqns \ref{single_chain_partition}. The exact form of the first line of this equation is somewhat arbitrary, and depends on which terms we extract from the prefactor of the partition function to make the logarithms dimensionless. However, we always subtract the free energy density of the homogeneous state with the same volume fraction of copolymer $A_\text{h}/Vk_\text{B}T$ (the first two lines of Eqn.\ \ref{SCFT_FE}) from $A/Vk_\text{B}T$, so this choice does not affect the final result.

In order to calculate the SCFT density profiles and free energy densities for a given volume fraction of copolymer, the set of simultaneous equations (\ref{SCFT_equations}) must be solved with the densities calculated as in Eqn.\ \ref{density}. To do this, we use a simple mixing iteration \cite{matsen2004}. First, we guess the form of the fields $w_i(\mathbf{r})$ and solve the diffusion equations \ref{diffusion} and \ref{copdiffusion} to calculate the propagators corresponding to these fields. From these, we calculate the densities using Eqns \ref{density}. New values for the fields are now calculated using the new $\phi_i(\mathbf{r})$. We then replace the $w_i(\mathbf{r})$ with a mixture of the old and new values of $w_i$ ($0.99w_i^\text{old}+0.01w_i^\text{new})$) and then recalculate the $\phi_i$. This approach proves more stable than simply replacing the old values of the $w_i$ with the new ones.

The procedure is repeated until the left and right hand sides of all the simultaneous equations (\ref{SCFT_equations}) differ by less than $10^{-5}$. For several systems, we have checked that the iteration arrives at the same solution for different initial $w_i$.

The diffusion equations are solved in spherically-symmetric, cylindrically-symmetric or planar geometries, depending on the structure we wish to study. For simplicity, we consider infinite cylinders and bilayers, allowing us to solve the SCFT equations in 1d rather than 2d. This means that we neglect the end cap energy of the cylinder and the edge energy of the bilayer. Since the aggregation number of cylinders and bilayers is typically much greater than that for spherical micelles, we expect this approximation to be good. 

In all cases, we impose reflecting boundary conditions at the origin and at the boundary of the system. A real-space finite difference algorithm \cite{num_rec} is used to solve the diffusion equations, in contrast to the Fourier space methods used in much of the SCFT literature \cite{cavallo}. A step size of $\Delta r= 4\,\text{\AA}$ is used for all geometries. It has been checked that decreasing the step size does not strongly change the $\phi_i(\mathbf{r})$ or the free energy densities. 

A key task in our calculation is to determine the micelle in each geometry which minimizes the total free energy of the system. Once we know the optimum micelle in each geometry, we can compare their free energies to find the morphology with the lowest free energy, i.e., the morphology that will be formed in a given blend. In our discussion of the SCFT method above, we considered a simple system of fixed volume and fixed copolymer volume fraction containing one micelle. To find the micelle of a given symmetry with the lowest free energy, we must consider how a system of many micelles minimizes its free energy. Consider a macroscopic copolymer/homopolymer blend whose copolymer volume fraction $\overline{\phi}_\text{PB}+\overline{\phi}_\text{PS}$, total volume $V_\text{T}$ and temperature $T$ are all fixed. The equilibrium state of this system can be found by minimizing the total free energy $F$, or equivalently the free energy density $F/V_\text{T}$ (since the total volume $V_\text{T}$ is constant). If the copolymer concentration is above a certain value (the {\em critical micelle concentration}), copolymer chains can either exist as monomers or in micelles. The number density of micelles is thus an internal degree of freedom and the macroscopic system varies this quantity (subject to the constraint of fixed copolymer volume fraction) in order to minimize the free energy density $F/V_\text{T}$. Explicit calculations on this many-micelle macrosopic system are extremely time-consuming even using SCFT. However, we can reduce the problem to one involving only a single micelle if we neglect inter-micellar interactions and the translational entropy of the micelles. The former is applicable if the micellar solution is sufficiently dilute while the latter introduces a (small) correction term to the free energy which will be included by hand later in this section. In this case, we can reduce the many-micelle system to a one-micelle system of volume $V$ and copolymer volume fraction $\overline{\phi}_\text{PB}+\overline{\phi}_\text{PS}$, where $V$ corresponds to the volume per micelle. We can then effectively vary the number density of micelles by varying $V$. If the free energy of this subsystem is $A$, we can then find the equilibrium state of the whole system by minimizing the free energy density $a=A/V$ with respect to $V$. Since each subsystem contains only one micelle, this procedure automatically yields the optimum micelle for a given geometry; that is the micelle with the lowest free energy per chain \cite{safran_book}.

We can show more formally that minimizing $A/V$ is equivalent to minimizing the total free energy $F$ of a system of $N$ micelles by writing
\begin{equation}
F(\overline{\phi}_\text{PB}+\overline{\phi}_\text{PS},V_\text{T})=NA(\overline{\phi}_\text{PB}+\overline{\phi}_\text{PS},V)
\label{full_system_FE}
\end{equation}
since all the subsystems are equivalent and contain the same volume fraction of copolymer as the whole system. We now wish to minimize this free energy subject to the constraint that the total volume of the system is conserved; that is, $NV=V_\text{T}$. The number of micelles and the volume of each subsystem are allowed to vary. Carrying out the constrained minimisation of Eqn.\ \ref{full_system_FE} using Lagrange multipliers, we find, as above, that minimizing the free energy of the whole system with the above constraint corresponds to minimizing the free energy {\em density} $a=A(\overline{\phi}_\text{PB}+\overline{\phi}_\text{PS},V)/V$ of the subsystem with respect to the subsystem volume $V$.

To our knowledge, this method of varying the size of the calculation box containing a single micelle in order to obtain information on a system of many micelles has not been used before: in earlier work, the box size is fixed \cite{vanlent}. A clear advantage of our approach is that it yields a well-defined value for the volume occupied by a micelle. This allows us to take into account the translational entropy of spherical micelles \cite{mayes_delacruz}. An estimate of the translational entropy per micelle can be obtained from a simple lattice model where the system is divided into cells of the volume of a single micelle. We adapt results from scaling theory studies of micelle formation \cite{leibler,mayes_delacruz} and find that the translational entropy per micelle is
\begin{equation}
S_\text{trans}=-k_\text{B}\left[\ln \left(\frac{V_\text{m}}{V}\right)+\left(\frac{V-V_\text{m}}{V_\text{m}}\right)\ln\left(\frac{V-V_\text{m}}{V}\right)\right]
\label{trans_ent}
\end{equation}
where $V_\text{m}$ is the volume of the micelle and $V$ is the volume of the subsystem containing the micelle. Note that the lattice model leading to Equation \ref{trans_ent} implicitly assumes that micelles are impenetrable. Equation \ref{trans_ent} thus also partially corrects for inter-micellar interactions which were neglected in the preceding discussion.

To estimate the micelle volume, we follow the same approach as in our earlier study of the radii of spherical micelles \cite{gbm}. This requires working definitions of the core radius $R_\text{c}$ and the corona thickness $L_\text{c}$. We define the core radius as that at which the local volume fractions of the core species PB and the corona species PS are equal: $\phi_\text{PB}(\mathbf{r})=\phi_\text{PS}(\mathbf{r})$. This choice is arbitrary; however, the boundaries are quite sharp and so the differences between different definitions are rather small.

To estimate $L_\text{c}$, we first calculate the radius of gyration of the corona from \cite{burchard}
\begin{equation}
R_\text{g}^2=\frac{\int r^2 (\phi_\text{PS}(r)-\phi_\text{PS}^\text{b})4\pi r^2\mathrm{d}r}{\int(\phi_\text{PS}(r)-\phi_\text{PS}^\text{b})4\pi r^2\mathrm{d}r}
\label{R_g_def}
\end{equation}
where $\phi_\text{PS}^\text{b}$ is the polystyrene concentration at the boundary of the subsystem. This bulk value must be removed to isolate the corona. We now calculate the thickness $L_\text{c}$ of the spherical shell with inner radius $R_\text{c}$ (the core radius as calculated above) which has the same radius of gyration as the corona. This is taken as an estimate of the corona thickness. $L_\text{c}$ is related to $R_\text{c}$ and $R_\text{g}$ by \cite{burchard}
\begin{equation}
R_\text{g}^2=\frac{3}{5}\frac{(R_\text{c}+L_\text{c})^5-R_\text{c}^5}{(R_\text{c}^3+L_\text{c}^3)-R_\text{c}^3}
\label{R_L_R}
\end{equation}
and can be determined numerically. The volume of the micelle can then be calculated directly from $R_\text{c}+L_\text{c}$. For cylindrical micelles and lamellar bilayers, the contribution of translational entropy to the free energy density vanishes since we consider infinite cylinders and bilayers.

We now have all the necessary techniques to calculate the optimum micelle of a given geometry. To begin, we perform an SCFT calculation at fixed subsystem volume, giving the density profile of a micelle and the free energy density of the subsystem. We then adjust the subsystem volume. This is achieved by changing the number of points on the grid on which we solve the diffusion equations whilst keeping the grid stepsize constant. This is repeated until we have located the minimum of the free energy density for the geometry under consideration. The free energy densities for the different shapes of aggregate are then compared to find which is most likely to form for a given blend.

Although having to minimize $a$ with respect to $V$ for each system parameter adds to the numerical burden of the calculation, the advantage of this method is that, by minimizing the free energy density, we avoid the awkward problem of trying to define the free energy per chain in the micelle, which is the basic quantity in simple theories of micellization \cite{safran_book}. Taking this latter approach would involve making ad hoc definitions concerning the boundary of the micelle in what is essentially a continuum calculation.

The only problem with our method is that, while the free energy of the micelle is modeled accurately by SCFT, the free energy of the bulk is modeled poorly \cite{cavallo, wijmans_linse, gbm}. Fortunately, we expect the error in calculating the free energy density introduced by the bulk to be small since most block copolymer chains in our calculations are in the micelle. The phase boundaries between different micelle shapes should therefore be fairly accurate. This will be confirmed by the work presented here. Nevertheless, the difficulty in modeling the bulk means that minimizing the free energy density does not necessarily yield the {\em density profile} of the optimum micelle (the micelle with the lowest free energy per chain). The reason for this is that, since the bulk copolymer concentration is severely underestimated \cite{cavallo, wijmans_linse, gbm}, fixing the overall copolymer volume fraction to a realistic value means that the number of chains in the micelle will be overestimated. In consequence, the profiles of spherical micelles obtained from our current minimization method are swollen compared to experimental data \cite{kinning}.

If we wish to study the density profiles of micelles {\em of a given geometry}, we must accept the fact that SCFT systematically underestimates the copolymer concentration with which the micelle coexists, and take an alternative approach to finding the optimum micelle. We have carried out this task elsewhere \cite{gbm}. Our method was to minimize the free energy per copolymer chain {\em in the micelle} \cite{safran_book}. We performed calculations on a single micelle in a system of {\em fixed} volume that was sufficiently large for the copolymer concentration to have attained a constant (bulk) value at the edge of the system. We then noted that this bulk copolymer concentration coexisting with the micelle is a monotonically increasing function of the bulk chemical potential. However, at equilibrium, the bulk chemical potential is equal to the free energy per chain in the micelle \cite{safran_book}. We can therefore minimize the free energy per chain in the micelle (as required) by minimizing the copolymer concentration at the edge of our system with respect to the overall copolymer concentration. In other words, we used the bulk copolymer concentration as a proxy for the free energy per chain in this approach. This avoids the ambiguities inherent in defining a free energy per chain in a continuum model.

This approach was successful in making quantitative predictions for the core radii for spherical copolymer micelles in homopolymer. However, it is not suitable for the current problem of studying micelle phase transitions, since the free energy per chain in the micelle may not be a monotonic function of bulk block copolymer concentrations across different micelle geometries. Therefore, to determine the phase boundaries between these different morphologies, the best method is to minimize the free energy density with respect to the subsystem volume $V$, as is done in the current paper. Once the micelle shape for a given system has been determined, more refined predictions for the micelle profile may be obtained, if required, by minimizing bulk block copolymer concentrations.

\section{Results and discussion}\label{results}

\begin{figure}
\includegraphics[width=\linewidth]{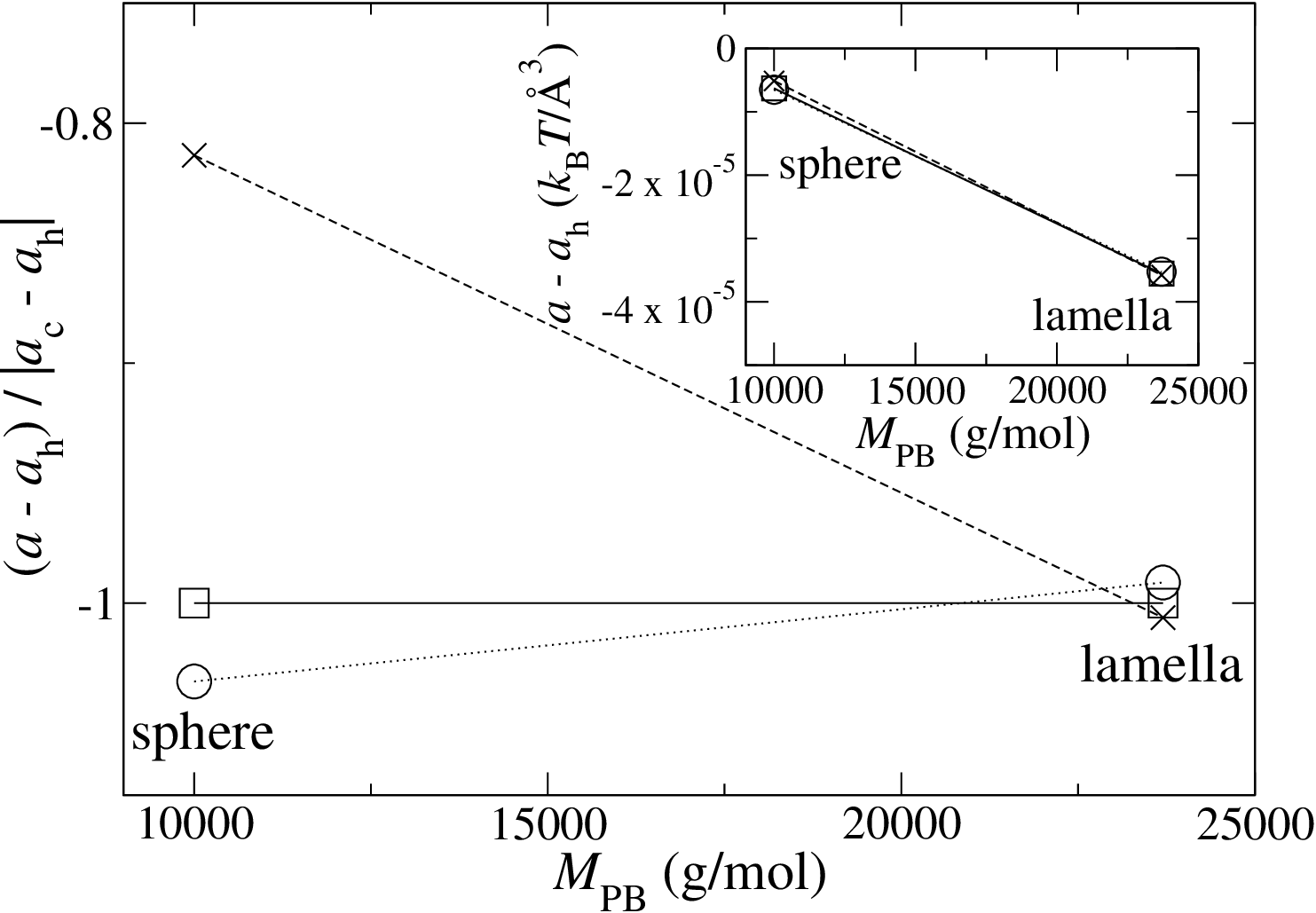}
\caption{\label{SB10-NN-3900}
}
\end{figure}

We begin by examining the shape transition that occurs as we move from a symmetric copolymer to one with a heavier polybutadiene core block at constant homopolymer molecular weight, and compare our predictions with the experimental results of Kinning, Winey and Thomas \cite{kinning_winey_thomas}. Our predictions are presented as follows. In the inset to Fig.\ \ref{SB10-NN-3900}, we plot the free energy density $a$ (minus the free energy density $a_\text{h}$ of the homogeneous state with the same copolymer weight percentage) for spheres, cylinders and bilayers of SB 10/10 and SB10/23 copolymers blended with 3900PS. However, it may quickly be seen that the differences in free energy densities between the different blends are much larger than those between the different morphologies in a given blend. This makes it difficult to see which free energy is the lowest for a particular blend and hence which shape is the most likely to form. To avoid this problem, we normalize our results by the magnitude of $a-a_\text{h}$ for the cylindrical morphology and plot the quantity $(a-a_\text{h})/|a_\text{c}-a_\text{h}|$. In the case of the cylinder, this is simply a horizontal line at $(a-a_\text{h})/|a_\text{c}-a_\text{h}|=-1$. The corresponding lines for the bilayer and sphere approach this from above and below respectively as the homopolymer molecular weight is increased, and the transitions between the morphologies may be clearly seen. We plot the data in this manner throughout the paper.

Firstly, we consider a blend of symmetric SB 10/10 copolymer with 3900PS. The weight percentage of copolymer is $13.0\,\%$. This system is found experimentally \cite{kinning_winey_thomas} to form spherical micelles. The reason for this is that the entropy of mixing between the small homopolymer 3900PS and the polystyrene blocks is very high. The homopolystyrene then swells the polystyrene corona, making the interface between the two species naturally very curved and causing spherical micelles to form. Our SCFT calculations (Fig.\ \ref{SB10-NN-3900}) also find that the sphere is the most favorable aggregate, with the lowest free energy density.

If the weight of the polybutadiene block is increased, a different morphology is found. Specifically, the copolymer is changed from SB 10/10 to SB 10/23. The same homopolymer 3900PS is still used, and the weight percentage of copolymer is $18.4\,\%$. Kinning, Winey and Thomas \cite{kinning_winey_thomas} found that this system formed multilamellar vesicles: concentric shells of copolymer. In line with this experimental finding, we predict that the bilayer is the most energetically favorable structure (see Fig.\ \ref{SB10-NN-3900}).

\begin{figure}
\includegraphics[width=\linewidth]{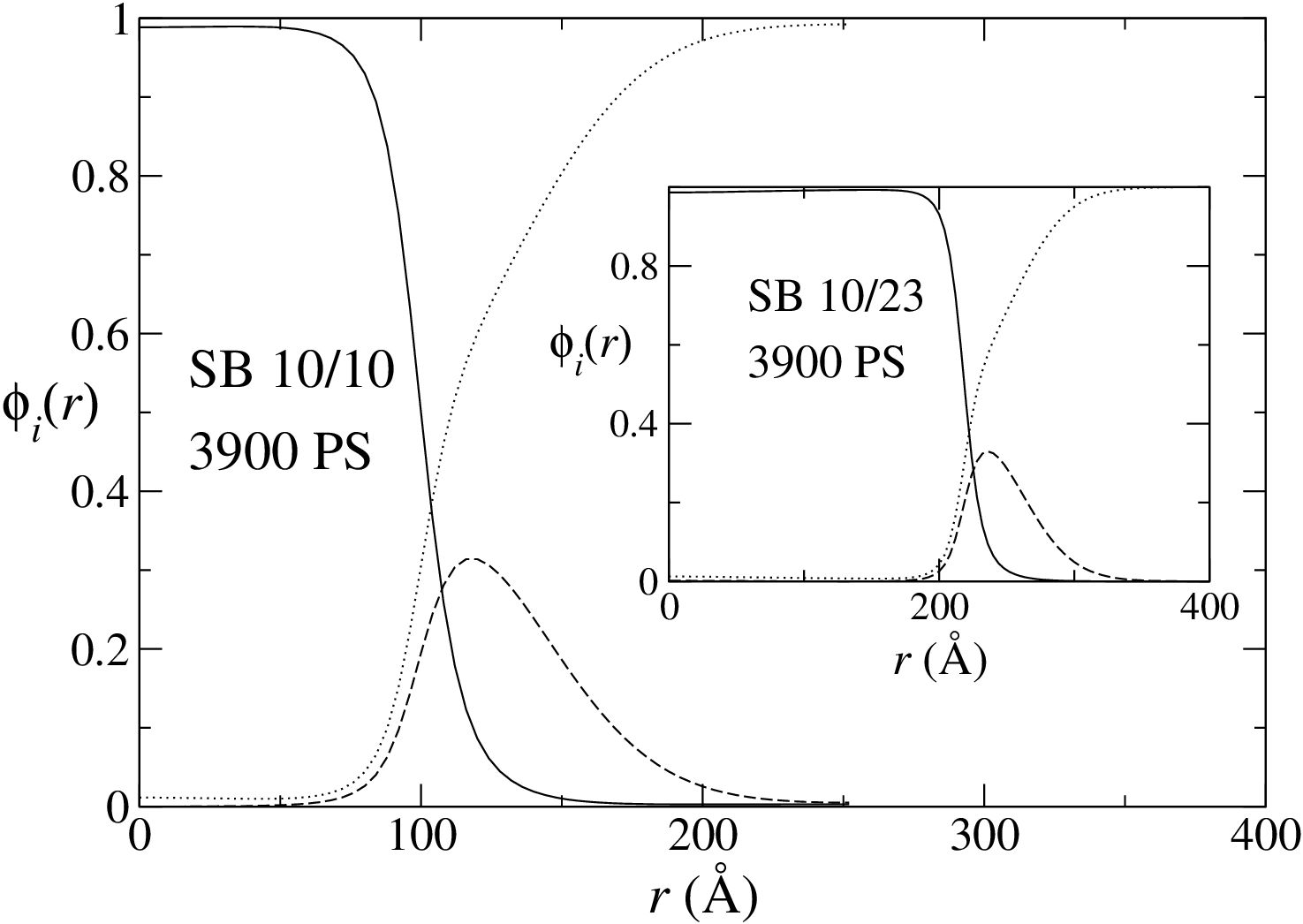}
\caption{\label{SB10-NN-3900phi}
}
\end{figure}

To gain physical insight into this transition, we plot cross-sections through two different spherical micelles in Fig.\ \ref{SB10-NN-3900phi}. In the main panel of Fig.\ \ref{SB10-NN-3900phi}, we plot the volume fraction profiles of the three different species for a spherical micelle in the first blend considered in Fig.\ \ref{SB10-NN-3900}: $13.0\,\%$ SB 10/10 in 3900 PS homopolymer. This structure has the lowest free energy density and hence is the most likely to form. The reason for this is that it has a corona that is strongly swollen by homopolymer and a relatively small core. This means that the interface between PS and PB is highly curved and the spherical micelle shown here is observed in experiment \cite{kinning_winey_thomas}.  

In the inset to Fig.\ \ref{SB10-NN-3900phi}, we show the volume fraction profiles for a spherical micelle in the second blend shown in Fig.\ \ref{SB10-NN-3900}, namely $18.4\,\text{wt}\,\%$ SB 10/23 copolymer in 3900 PS. The spherical micelle plotted here is predicted to be the {\em least} energetically favorable structure, and the physical reasons are clear from its density profile. Here, the corona is also clearly swollen by the small homopolymer, and the volume fraction profiles for the corona and homopolymer are very similar to those seen in the blend containing the symmetric copolymer SB 10/10 (main panel of Fig.\ \ref{SB10-NN-3900phi}). However, the polybutadiene block is much heavier in SB 10/23, and the radius of the core is hence much larger. This compensates for the swelling of the corona, and the curvature of the PS/PB interface is much smaller. As a result, bilayers are formed, which, in the experimental system, wrap up into multilamellar vesicles to avoid energy penalties due to the formation of edges. The spherical micelle shown here has a higher free energy density than either of the other two morphologies and is not found experimentally \cite{kinning_winey_thomas}.

\begin{figure}
\includegraphics[width=\linewidth]{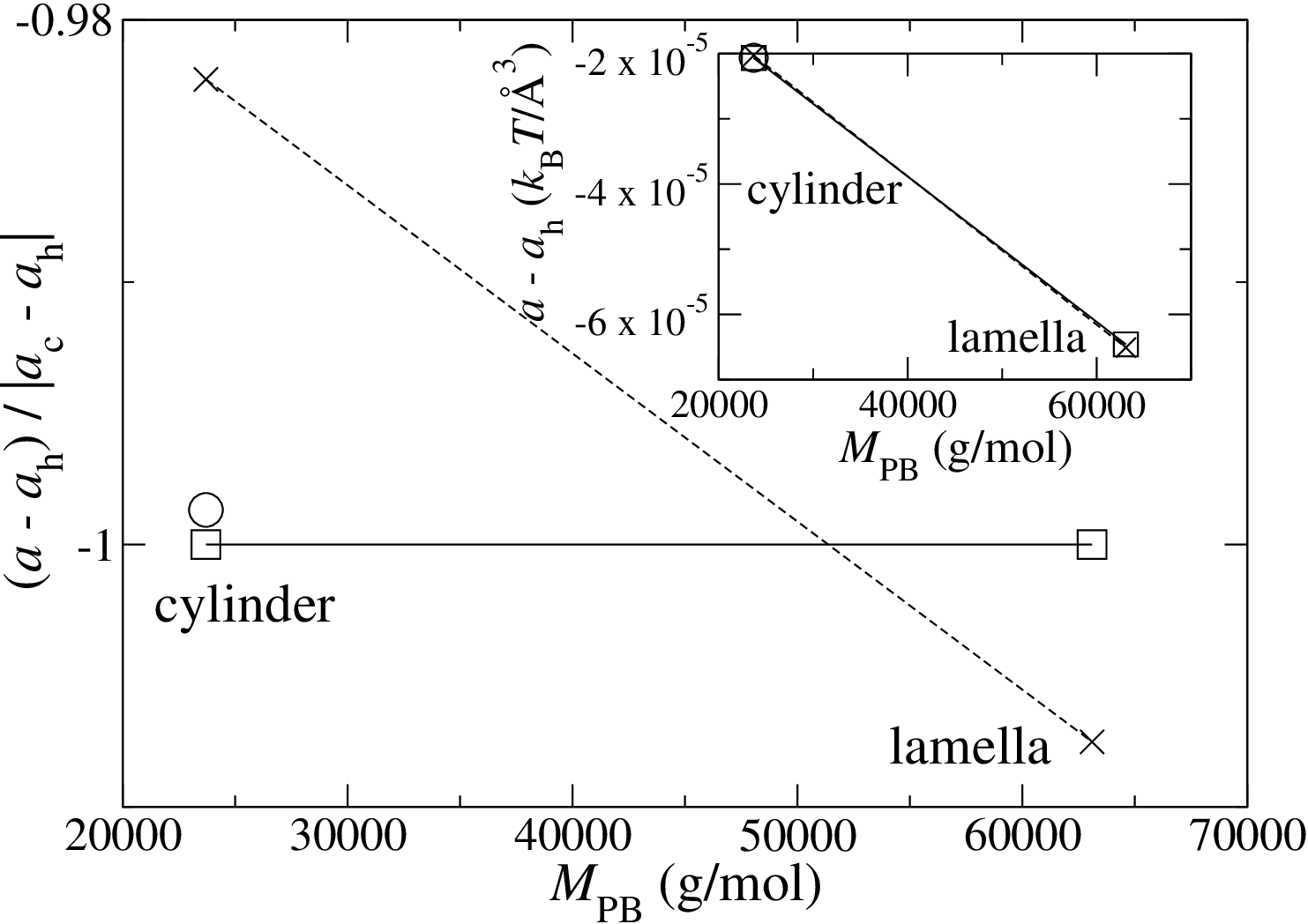}
\caption{\label{SB10-NN-2100}
}
\end{figure}

A similar shape transition may be seen if we consider blends of asymmetric copolymer with the very short homopolymer 2100 PS. If the SB 10/23 copolymer discussed above is blended with 2100 PS rather than 3900 PS, scattering and TEM experiments (at copolymer weight percentage $17.8\,\%$) \cite{kinning_winey_thomas} find that it forms spherical micelles instead of multilamellar vesicles. As above, the reason for this is that the entropy of mixing between the polystyrene corona blocks and the small homopolymer is very high. The corona thus becomes very swollen, outweighing the effect of the relatively large core. This results in a highly curved PS/PB interface and spherical micelles are observed. Our calculations (Fig.\ \ref{SB10-NN-2100}) predict that the cylinder is slightly more favorable than the sphere seen in experiments \cite{kinning_winey_thomas}; however, both structures have clearly much lower free energy densities than the bilayer. In addition, the experiments necessarily work with a limited selection of homopolymers, and there will be occasions when a relatively small error in the calculation of the free energy causes a morphology to be incorrectly predicted. We can also understand the overestimation of the favorability of the cylindrical micelle by recalling that, by considering an infinite cylinder, we neglect the free energy penalties due to endcaps and the curvature of the micelle.

In electron micrographs of a blend of the highly asymmetric copolymer SB 10/65 with 2100 PS at $13\,\text{wt}\,\%$, a lamellar structure is seen \cite{kinning_winey_thomas}. This is also predicted by self-consistent field theory (see Fig.\ \ref{SB10-NN-2100}), which finds that the lamella has a much lower free energy density than the cylinder. Furthermore, in this blend, we were unable to find a minimum of the free energy density corresponding to a spherical micelle and therefore predict that this structure is unstable.

\begin{figure}
\includegraphics[width=\linewidth]{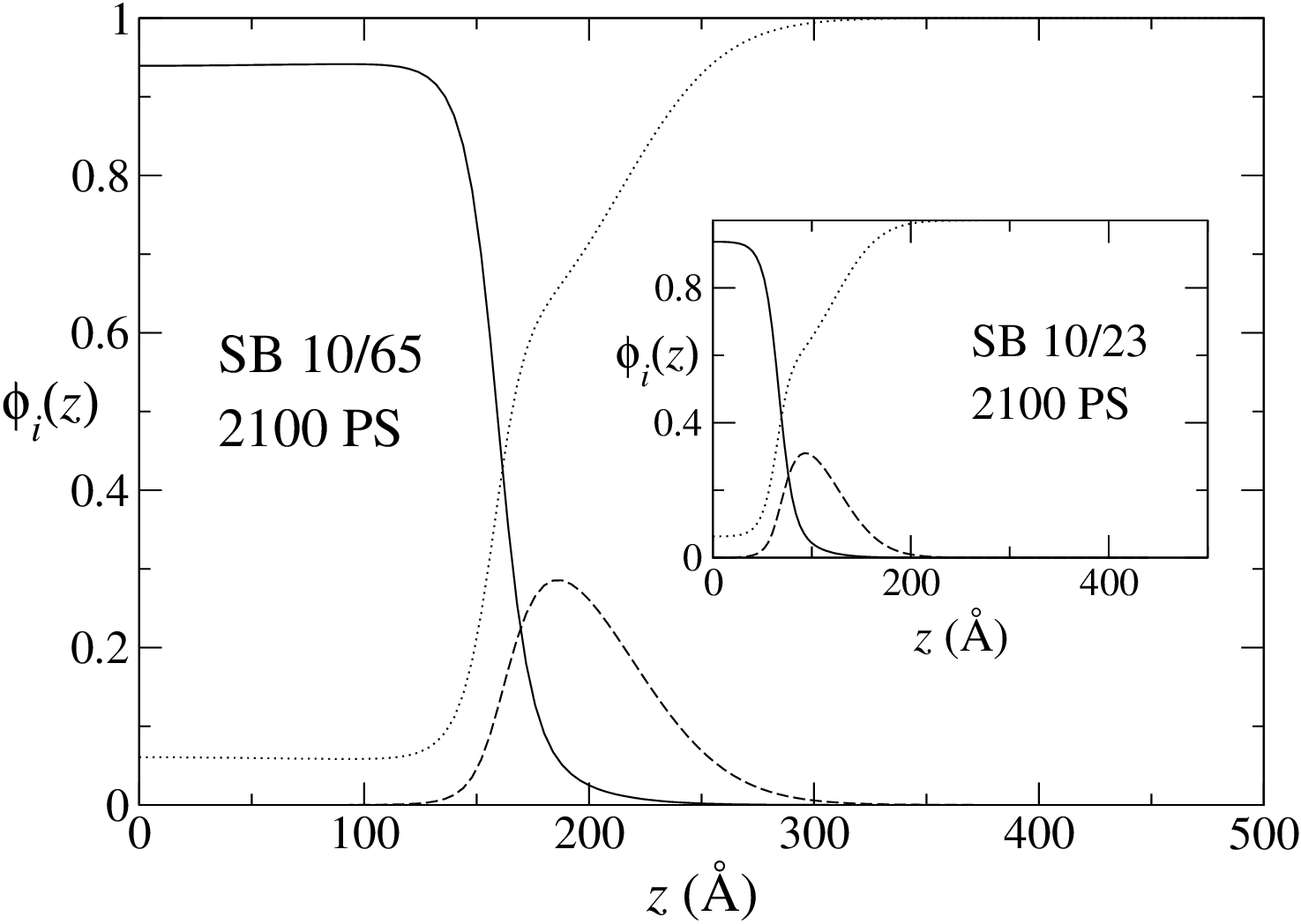}
\caption{\label{SB10-NN-2100phi}
}
\end{figure}

Again, plotting the volume fraction profiles of favorable and unfavorable aggregates allows us to illustrate the physical principles behind the shape transition. Here, we focus on the bilayer. In the blend containing SB 10/65, this structure has the lowest free energy density and is observed in experiment \cite{kinning_winey_thomas}. We plot its cross-section in the main panel of Fig.\ \ref{SB10-NN-2100phi}. The polystyrene blocks of the copolymer are seen to be significantly swollen by the small homopolymer. However, the large core block compensates for this, reducing the curvature of the interface and leading to the formation of the planar structure. This is in contrast to the system with SB 10/23 copolymer shown in the inset to Fig.\ \ref{SB10-NN-2100phi}. Here, the swelling of the polystyrene blocks is also high, and the density profiles of the corona and homopolystyrene are very similar to those seen in the SB 10/65 blend. However, in this system, the polybutadiene blocks are much shorter, and the swelling of the polystyrene means that the PS/PB interface naturally has a higher curvature. The bilayer shown here therefore has a much higher free energy density than those of the sphere and cylinder and is hence not seen in experiment \cite{kinning_winey_thomas}. It is also interesting to note that, in both cases, the short 2100 PS homopolymer is predicted to penetrate significantly into the PB core (see Fig.\ \ref{SB10-NN-2100phi}). However, the degree of penetration is similar for both copolymers and the shape transition is not driven by this effect but by the competition between the length of the core PB blocks and the swelling of the PS corona.

Next, we study the shape transitions that may be induced by varying the homopolymer molecular weight in a blend of symmetric poly(styrene-butadiene) diblocks and homopolystyrene at constant copolymer weight percentage. We consider four samples, all studied experimentally by Kinning, Winey and Thomas \cite{kinning_winey_thomas}. In all cases, $12.5\,\%$ weight percentage of the symmetric copolymer SB 20/20 is blended with homopolymer. However, the homopolymer molecular weight is increased from blend to blend. For the lightest three homopolymers (2100PS, 3900PS and 7400PS), the experiments find spherical micelles. Like the blend of SB10/10 with 3900PS (see Fig.\ \ref{SB10-NN-3900}), these blends consist of a symmetric copolymer blended with a relatively light homopolymer, which swells the corona blocks and leads to a highly curved PS/PB interface. For the heaviest homopolymer (17000PS), the swelling of the corona is less pronounced, the interface between PS and PB is less curved, and cylindrical micelles are observed.

\begin{figure}
\includegraphics[width=\linewidth]{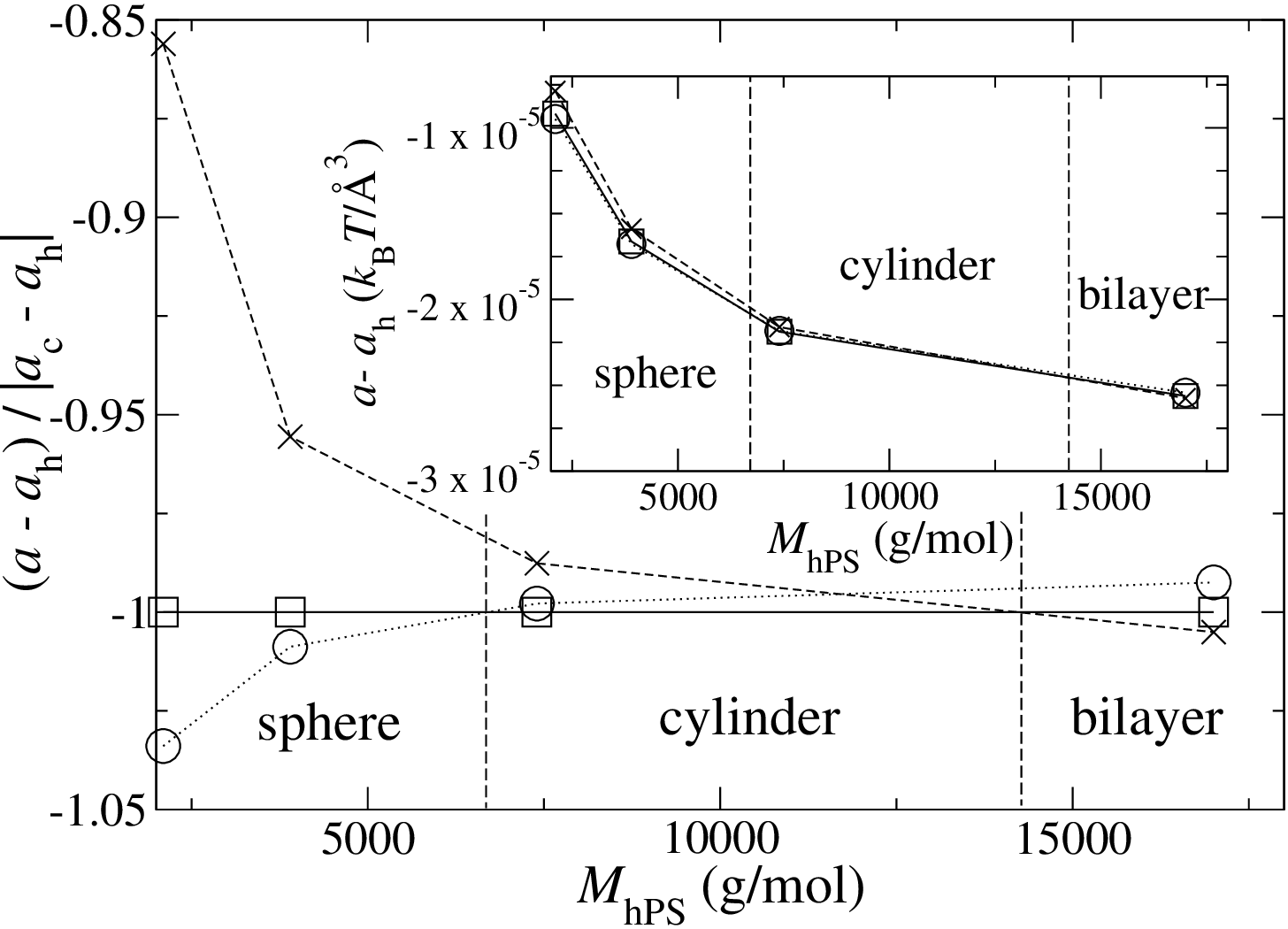}
\caption{\label{SB20-20}
}
\end{figure}

This trend away from spherical micelles as the homopolymer weight is increased is reproduced by SCFT (see Fig.\ \ref{SB20-20}). For the first two blends, we find, in line with experiment \cite{kinning_winey_thomas}, that the spherical micelle has the lowest free energy density and hence is most likely to form. In the case of the next blend (7400PS), we predict that the cylinder has a slightly lower free energy density than the sphere. As stated above, the experiments find that this blend forms spherical micelles \cite{kinning_winey_thomas}; however, we find that the difference between the free energy densities of the spherical and cylindrical morphologies is rather small and both micelles are clearly more favorable than the bilayer. As discussed above, the energetic favorability of the cylindrical micelle with respect to the spherical micelle is slightly overestimated probably due to our neglect of free energy penalties due to the endcaps of the cylinder.

For the blend with the heaviest homopolymer (17000PS), we predict that the bilayer has the lowest free energy density (see Fig.\ \ref{SB20-20}), whilst the experiments find cylindrical micelles. We believe that this discrepancy arises because, by considering ideal infinite structures, we have neglected edge effects that make the heavier of the two competing structures (here, the bilayer) less likely to form. Specifically, for micelles with aggregation number $N$, the edge penalty per copolymer falls off as $\sim N^{-1}$ for cylinders but only as $\sim N^{-1/2}$ for bilayers \cite{safran_book}. We therefore expect that neglecting edge effects overestimates the stability of bilayers relative to cylinders.

\begin{figure}
\includegraphics[width=\linewidth]{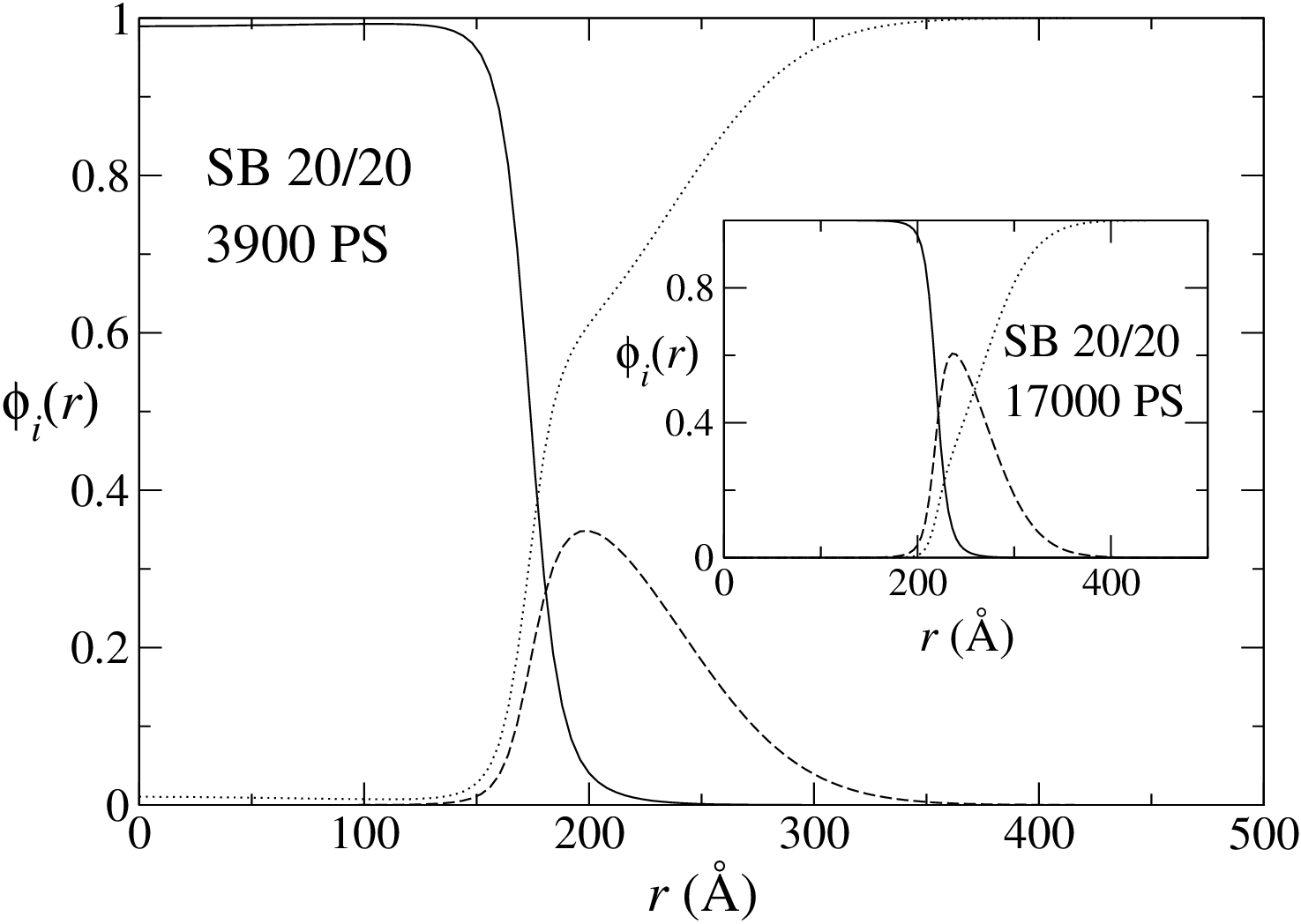}
\caption{\label{SB20-20phi}
}
\end{figure}

To illustrate the swelling of the polystyrene corona by homopolystyrene that drives the transitions between the different morphologies, we proceed as before and plot cross-sections through two different spherical micelles in Fig.\ \ref{SB20-20phi}. The main part of the figure shows the volume fraction profiles for the core, corona and solvent for a blend of SB 20/20 with 3900PS with $12.5\,\%$ copolymer by weight. This is the second system plotted in Fig.\ \ref{SB20-20}, and forms spherical micelles. It can immediately be seen that the corona is strongly swollen by the (small) homopolymer: the maximum local volume fraction of the polystyrene blocks is around $0.35$. To compensate for the swelling of the corona \cite{kinning}, the core radius is relatively small: around $180\,\text{\AA}$. In consequence, the interface between PB and PS is naturally strongly curved, and the spherical micelle shown here is the most favorable structure.

This is in contrast to the system shown in the inset to Fig.\ \ref{SB20-20phi}: a blend of SB 20/20 copolymer with 17000PS, again at copolymer weight percentage $12.5\,\%$. The spherical micelle plotted here is predicted to be the least energetically favorable structure. The reasons for this are as follows. Firstly, the heavy 17000PS homopolymer mixes much less well with the polystyrene blocks, so that the corona is much less swollen by homopolymer. This can be seen from the fact that the peak local volume fraction of the polystyrene is far higher (around $0.57$) than in the sphere-forming blend shown in the main panel of Fig.\ \ref{SB20-20phi}. Secondly, the core radius is approximately $220\,\text{\AA}$: significantly higher than in the blend with 3900PS homopolymer. The interface between the polybutadiene core blocks and the polystyrene corona blocks is therefore naturally less curved, and the spherical micelle does not form.

We summarize our results in Table \ref{tab:summary}, listing the experimentally- and theoretically-determined morphologies for all the blends discussed above. All trends are correctly predicted and the correct shape is predicted for five out of the eight blends. The percentage difference between the free energies (relative to the homogeneous state) of the shape predicted to be the most favorable and that predicted to be the next most favorable is also included for each blend. Note that the three samples for which the shape is incorrectly predicted (SB 10/23 in 2100 PS, SB 20/20 in 7400 PS and SB 20/20 in 17000 PS) also have the three smallest free energy differences between the two most favorable states. This suggests that, in these cases, we are close to the transitions between morphologies and any slight inaccuracy in the theory or in the measurement of experimental parameters could lead to an incorrect prediction of the shape. We emphasize that our calculations contain no adjustable parameters: all the required input concerning the polymer properties (such as the interaction energy density $\Lambda$) has been determined from experiments that do not involve micelle formation. Given this fact, we believe that the agreement between theory and experiment is excellent.

\section{Conclusions}\label{conclusions}

We have found that self-consistent field theory gives a good description of the different isolated structures that form in a blend of diblock copolymers and homopolymer. In the majority of cases, SCFT predicts the morphology seen in experiment \cite{kinning_winey_thomas}, and all qualitative effects (such as the trend away from spherical micelles as the homopolymer weight is increased) are reproduced.

Even when the shape is not correctly predicted, the difference between the free energy density of the structure seen in experiment and that predicted to have the lowest free energy density by SCFT is very small (see, for example, the first set of points in Fig.\ \ref{SB10-NN-2100}, where SCFT predicts a cylindrical structure but spheres are seen in experiment). The tendency of our SCFT calculations to overestimate the favorability of heavier aggregates may be understood by considering finite size contributions to the free energy, such as the energy penalties due to the cylinder endcaps and bilayer edges, that are neglected by our approach.

In summary, we have shown that SCFT provides a very good description of micelle shape transitions and hence that it is a suitable tool for the study of isolated block copolymer aggregates, provided the limitations of the theory are recognized when identifying the appropriate free energy to minimize. In particular, the free energy density must be minimized with respect to the volume of the subsystem containing a micelle to calculate shape transitions, whilst the free energy per chain in the micelle (in practice, the bulk block copolymer concentration) should be minimized to give more refined predictions of the micelle density profile.

This work was supported by the UK Technology Strategy Board, Unilever and ICI. M.J.G. thanks R. M. L. Evans for useful discussions.

\bibliographystyle{apsrev}

\newpage
\begin{table}
\begin{center}
\begin{tabular}{|x{1in}|x{1in}|x{1in}|x{1in}|x{1in}|x{1in}|}
\hline
SB & hPS & wt \% & experiment & theory & $(a_2-a_1)/|a_1|\,\%$ \tabularnewline \hline
SB 10/10 & 3900PS & 13.0 & S & S & 3.2 \tabularnewline \hline
SB 10/23 & 3900PS & 18.4 & L & L & 0.6 \tabularnewline \hline
SB 10/23 & 2100PS & 17.8 & S & C & 0.1 \tabularnewline \hline
SB 10/65 & 2100PS & 13.0 & L & L & 0.7 \tabularnewline \hline
SB 20/20 & 2100PS & 12.5 & S & S & 3.3 \tabularnewline \hline
SB 20/20 & 3900PS & 12.5 & S & S & 0.9 \tabularnewline \hline
SB 20/20 & 7400PS & 12.5 & S & C & 0.2 \tabularnewline \hline
SB 20/20 & 17000PS & 12.5 & C & L & 0.5 \tabularnewline \hline
\end{tabular}
\end{center}
\caption{\label{tab:summary} Summary table comparing experimental results with our theoretical predictions. The first column lists the copolymer sample, the second column the homopolymer with which it is blended, and the third column the copolymer weight percentage. The fourth and fifth columns show the experimental results and theoretical predictions respectively: S stands for sphere, C for cylinder and L for lamella. The sixth column lists the percentage difference between the free energies (relative to the homogeneous state) of the shape predicted to be the most favorable and that predicted to be the next most favorable. Note that the three samples for which the shape is incorrectly predicted SB 10/23 in 2100 PS, SB 20/20 in 7400 PS and SB 20/20 in 17000 PS) also have the three smallest free energy differences between the two most favorable states. This suggests that, in these cases, we are close to the transitions between morphologies and any slight inaccuracy in the theory or in the measurement of experimental parameters could lead to an incorrect prediction of the shape.}
\end{table}
\newpage
\section*{List of figures}\label{figlist}
\begin{enumerate}
\item The inset shows the free energy density (minus that of the corresponding homogeneous state) predicted by SCFT for two blends: one of the symmetric copolymer SB 10/10 with the homopolymer 3900 PS; the other of the asymmetric copolymer SB 10/23 with the same homopolymer. The free energy density is plotted against the polybutadiene block molecular weight. The weight percentage of copolymer is $13.0\,\%$ in the first blend and $18.4\,\%$ in the second.  The main panel shows the same data normalized with respect to the magnitude of the free energy density of a cylindrical micelle. The data for the spherical micelle are plotted as circles connected by dotted lines. Those for the cylinder are plotted as squares linked by full lines, and those for the lamella are shown as crossed connected by dashed lines.
\item The main panel shows the volume fraction profiles for the spherical micelle that forms in a blend of symmetric SB 10/10 copolymer with 3900 PS homopolystyrene. This micelle is predicted to have the lowest free energy density of the three morphologies and is observed in scattering and TEM experiments. The inset shows the corresponding plots for a spherical micelle in a blend of the asymmetric copolymer SB 10/23 with the same homopolymer. This structure is not energetically favorable and is not seen in experiment. The reason for this can be seen by comparing the core sizes in the two blends. In the blend with the symmetric copolymer SB 10/10, the core radius is relatively small due to the short core blocks. Together with the highly swollen corona, this leads the interface between PS and PB to be strongly curved and spherical micelles to form. In the asymmetric blend (SB 10/23), the core is much larger due to the heavier PB core blocks. This compensates for the swelling of the corona and reduces the curvature of the PS/PB interface. The spherical micelle shown here thus has a high free energy density and is not observed in experiment.
\item The inset shows the free energy density (minus that of the corresponding homogeneous state) predicted by SCFT for two blends: one of the moderately asymmetric copolymer SB 10/23 with the homopolymer 2100 PS; the other of the highly asymmetric copolymer SB 10/65 with the same homopolymer. The free energy density is plotted against the polybutadiene block molecular weight. The weight percentage of copolymer is $17.8\,\%$ in the first blend and $13.0\,\%$ in the second.  The main panel shows the same data normalized with respect to the magnitude of the free energy density of a cylindrical micelle. The data for the spherical micelle are plotted as circles connected by dotted lines. Those for the cylinder are plotted as squares linked by full lines, and those for the lamella are shown as crossed connected by dashed lines. Note that we found no free energy minimum corresponding to the spherical micelle for the SB 10/65 blend: this structure is thus predicted to be unstable.
\item The main panel shows the volume fraction profiles for the flat bilayer that forms in a blend of the highly asymmetric SB 10/65 copolymer with 2100 PS homopolystyrene. This structure is predicted to have the lowest free energy density of the three morphologies and is observed in scattering and TEM experiments. The inset shows the corresponding plots for a bilayer in a blend of the moderately asymmetric copolymer SB 10/23 with the same homopolymer. This structure is not energetically favorable and is not seen in experiment. The reason for this can be seen by comparing the thicknesses of the PB layer in the two blends. In the blend with SB 10/65, the PB layer is thick due to the heavy core blocks. This balances the effect of the highly swollen corona, and leads the PS/PB interface to be flat and bilayers to form. In the less asymmetric blend (SB 10/23), the PB layer is much less thick due to the lighter core blocks. This thinner layer cannot compensate for the swelling of the corona and the PS/PB interface is much more curved. The flat bilayer shown here thus has a high free energy density and is not observed in experiment.
\item The inset shows the free energy density predicted by SCFT for blends of the symmetric copolymer SB 20/20 with a range of homopolymers. The free energy density is measured with respect to that of the homogeneous blend with the same weight fraction of copolymer and is plotted against homopolymer molecular weight. In all cases, the weight percentage of copolymer is $12.5\,\%$. The main panel shows the same data normalized with respect to the magnitude of the free energy density of a cylindrical micelle. The free energy density of this morphology then appears as a horizontal line, and the corresponding data for the lamella and sphere approach it from above and below. The data for the spherical micelle are plotted as circles connected by dotted lines. Those for the cylinder are plotted as squares linked by full lines, and those for the lamella are shown as crossed connected by dashed lines. The vertical dashed lines mark the approximate boundaries between the different morphologies.
\item 
The main panel shows the volume fraction profiles for the spherical micelle that forms in a blend of symmetric SB 20/20 copolymer with 3900PS homopolystyrene. This micelle is predicted to have the lowest free energy density of the three morphologies and is observed in scattering and TEM experiments. The inset shows the corresponding plots for a spherical micelle in a blend of SB 20/20 with the much heavier homopolymer 17000 PS. This structure is not energetically favorable and is not seen in experiment. The reason for this can be seen by comparing the degree of swelling of the corona in the two blends. In the blend with the lighter homopolymer (3900 PS), the corona is strongly swollen by homopolymer. This leads to a highly curved interface between the core and corona and the formation of spherical micelles. In the other blend (17000 PS), the swelling of the corona is less pronounced, the natural curvature of the interface is less, and spherical micelles are not seen.
\end{enumerate}

\end{document}